\documentstyle[buckow]{article}

\def\beq{\begin{equation}}

\def\eeq{\end{equation}}

\begin {document}

%HU-EP-00/57

\def\titleline{
About The Enhan\c{c}on Mechanism
}

\def\authors{
Paolo Merlatti\1ad}

\def\addresses{
\1ad
Dipartimento di Fisica Teorica, Universit\`a di Torino\\
via P. Giuria 1, I-10125 Torino, Italy\\
{\tt Paolo.Merlatti@to.infn.it}
}

\def\abstracttext{

I study the enhan\c{c}on mechanism for fractional D-branes in conifold and orbifold
backgrounds and show how it can resolve the ``repulson'' singularity of these geometries. In
particular I show that the consistency of the excision process requires that the interior
space be not empty.}

\large

\makefront

\section{Introduction}

In this contribution I describe the general phenomenon of the appearance of naked singularities in those gravitational backgrounds that are dual to non-conformal gauge theories. In particular, I would like to point out that in many cases these singularities are actually removed by different mechanisms. What is most striking is that the mechanism of removing those singularities seems to be strictly related to the non perturbative dynamics of the strongly coupled dual gauge theory.

In theories that are ${ \cal{N} } =2$ supersymmetric the relevant mechanism is the so-called enhan\c{c}on \cite{pol}. It removes a family of time-like singularities from certain supergravity space-times by forming a shell of (massless) branes on which the exterior geometry terminates. When this happens, one can say that the singularity is ``excised'' so that the background becomes acceptable. However this is not the whole story. Indeed the geometry inside the shell has now to be completely determined, since the enhan\c{c}on is just a signal that the interior region must be replaced by new spacetime.

From the gauge theory point of view the enhan\c{c}on has a simple interpretation. Indeed the ``enhan\c{c}on locus'' corresponds to the surface, in the moduli space of the gauge theory, where the instanton effects become important and the perturbative description of the theory ceases to be accurate. Therefore the appearance of the enhan\c{c}on shell prevents one from finding a gravitational picture of the strongly coupled infrared gauge theory. I would like to stress that finding it would correspond to determine the internal geometry.

Let us now consider gauge theories that are ${ \cal{N} } =1$ supersymmetric and have a gravitational dual description. There is a remarkable example \cite{KT,KS} of such a theory, in which the proper gravitational dual background is flat-space `times' a conifold. In this case, the mechanism that removes the singularity is the deformation of the conifold. This mechanism gives a nice geometrical picture of the strong non-perturbative dynamics of the gauge theory. This is very different from the enhan\c{c}on, which is just a signal of the appearance of non-negligible infrared effects in the gauge theory that have no longer a dual gravitational description.  However, I will show that also in the conifold case there is a consistent non-singular solution with a configuration very similar to that of the enhan\c{c}on solution of the ${ \cal{N} } = 2$ case.

The scenario that emerges from this analysis is the following. As far as one considers the ultraviolet regime of gauge theories, one finds qualitative and quantitative agreement with their gravitational dual descriptions. When instead one reaches regimes where the non-perturbative corrections become important, the gravitational description becomes more obscure (the space-time exhibits naked singularities) and this fact is signaled by the enhan\c{c}on mechanism that excises the singularities. This excision process is unavoidable in view of some stringy phenomena, such as enhan\c{c}ed gauge symmetries and the vanishing of certain D-brane tensions. However, in some cases, one is able to find non singular solutions even without appealing to the enhan\c{c}on phenomenon. In those cases, we are able to explore the spacetime inside the shell, determining thus the interior geometry that properly describes the dual gauge theory.

%%%%%%%%%%%%%%%%%%%%%%
\section{The Enhan\c{c}on}
%%%%%%%%%%%%%%%%%%%%%%

The main purpose of this contribution is to analyze the enhan\c{c}on mechanism on the conifold. On this background, we lack of an explicit description, in terms of two-dimensional conformal field theory, of ordinary and fractional D$p$-branes;
therefore many techniques such as the boundary state formalism or the brane-probe computations, cannot be applied. Then this analysis has to be performed using exclusively supergravity tools. However, the enhan\c{c}on mechanism in the presence of fractional branes occurs also in the case of orbifold backgrounds \cite{pol,myers,bertolini,gauge,frau}. These are more tractable examples since we have an exact stringy description of these geometries. In these backgrounds we can compare the supergravity solutions with the insights we can have from the conformal description.
Indeed, it is possible to show \cite{frau,en} that the definition of the enhan\c{c}on locus as the locus where the metric on the moduli space of the gauge theory living on the world-volume of the D-branes vanishes, coincides with a purely gravitational definition. From a gravitational point of view, the enhan\c{c}on corresponds to a modification of the naive supergravity solution, where the constituent branes smear out from being point-like to being a ``sphere''. Moreover, it's possible to see that they are massless on this sphere. We can describe this gravitational phenomenon by means of `incision' computations \cite{pol}, finding perfect agreement with the stringy computations about the world-volume gauge theory.

The analysis of how supergravity results perfectly match with stringy computations in the orbifold case is made in \cite{en}. This analysis corroborates, in my opinion, the interpretation I propose also for the conifold backgrounds, where we can do only gravitational calculations. Let us now study the enhan\c{c}on mechanism on the conifold.

%%%%%%%%%%%%%%%%%%%%%%%%%%%%%%%%%%%%%%%%%%%%%%%%%%%%%%%%%%%%%%%%%%%%%%%%%%%

%%%%%%%%%%%%%%%%%%%%%%%%%%%%%%%%%%%%%%%%%%%%%%%%%%%%%%%%%%%%%%%%%%%%%%%

\vskip0.5cm
\noindent{\bf The enhan\c{c}on on the conifold}

\noindent Type IIA and type IIB fractional D$p$-brane solutions in conifold backgrounds have been
extensively studied in \cite{herzog}. They are warped product of $I\! R^{1,p}$ flat
space-time directions and a Ricci flat, ($9-p$)-dimensional cone ${\mathcal{C}}_{9-p}$.
Since the brane are space filling, the warp-factor depends only on the radial coordinate
of the cone. Because the cone is Ricci flat, the base of the cone is an
($8-p$)-dimensional Einstein manifold $X_{8-p}$ with metric $h_{ij}$. It's assumed this
Einstein manifold has a harmonic 2-form $\omega_2$, so that wrapping a D$(p+2)$-brane
around the 2-cycle corresponding to $\omega_2$, and letting the remaining $p+1$ dimensions
fill $I\! R^{1,p}$, gives origin to a fractional D$p$-brane. For the sake of clarity, we describe now
the fractional D$0$-branes, but everything is valid for every $p$.

The fractional D$0$-brane solution is:
\begin{eqnarray}
\nonumber \mbox{e}^{\Phi} &=& H(r)^{3/4}  \\
ds_E^2 &=& g_s^{1/2}\ \Big[ -H^{-7/8}dt^2+H^{1/8}(dr^2\ +\ r^2h_{ij}dx^i\ dx^j)\Big]
\end{eqnarray}
The non-zero R-R field strengths are:
\begin{eqnarray}
\nonumber F_2  &=& dt\wedge dH^{-1}\\
\tilde{F}_4 &=& \frac{H^{-1}}{r^4}~Q~dt\wedge dr\wedge\omega_2
\end {eqnarray}
where $\tilde{F}_4\ =\ F_4- C_1\wedge H_3$, $Q\sim   g_sM$, $M$ is the number of fractional D$p$-branes
and $g_s$ is the string coupling constant. \\
We want now to express, for more generality, the NS-NS two form field ($B_2$) and the warp
factor $H$ for $p$ generic ($p<3$):
\begin{eqnarray}
\nonumber B_2 &=& \frac{Q}{r^{3-p}}\frac{\omega_2}{p-3}\\
\label{Hsol} H(r) &=& \frac{\rho}{r^{7-p}}\ -\ \frac{Q^2}{(3-p)(10-2p)r^{10-2p}}
\end{eqnarray}
where as usual $\rho\sim g_s N$ and $N$ is the number of ordinary D$p$-branes. The warp
factor has to satisfy the following equation of motion (it comes from that for the R-R
field strength $F_2$):
\begin{equation}
\label{H} \frac{d}{dr}\Big(r^{8-p}\frac{d}{dr}H(r)\Big)=-\frac{Q^2}{r^{4-p}}
\end{equation}

This is the Poisson equation for the $H$-potential. It has been integrated in (\ref{Hsol})
with the boundary condition that $H$ approaches zero as $r\to\infty$, so this solution is
valid for small $r$ (near-horizon limit) and consequently we are studying the gauge theory
(according to the gauge/gravity dual principle) in the infrared. \\ Note also that the
$r$-dependence of the source term seems to indicate a spatial extension of the source in
the transverse directions. Moreover, from the precise form of the dependence, it's
possible to argue that it spreads over ($4-p$) directions of the base of the cone.
\\ Moreover, as already pointed out by the authors of \cite{herzog}, such solutions possess a
naked singularity at $r=r_0$, in the IR. This singularity is of ``repulson'' type because
before reaching it, the derivative of $G_{00}$ changes sign at $r_*>r_0$ (where
$H'(r_*)=0$) and consequently the gravitational force changes its sign.

We want now study the possibility that an excision process takes place, as it has been
shown to happen in \cite{myers} in a different context. There, it has been studied this
mechanism for branes wrapped around the $K3$ surface, but we think that also in the
present case it happens something very similar, giving rise to the enhan\c{c}on. \\ Then,
in order to avoid the singularity, we can imagine that the D$p$-branes are not in the
origin, but they distribute on a surface at some $r=r_i\geq r_*$. For generic $p<3$ we can
try to calculate the stress-energy tensor of these source branes, interpreted as the
discontinuity in the extrinsic curvature (we are making the calculation in the Einstein
frame). 

If we now suppose that in the interior region ($r<r_i$) the space is flat, we find \cite{en} that the stress-energy tensor supported at the junction surface is:
\begin{eqnarray}
S_{\mu\nu} &=& \frac{1}{2\kappa^2}\frac{1}{\sqrt{G_{rr}}}\Big(\frac{H'}{H}\Big)G_{\mu\nu} \\
\nonumber S_{ij} &=& 0
\end{eqnarray}
According to the second line, we see that there is no stress in the directions transverse
to the branes, as it has to be since the branes in consideration are BPS.\\
From the first line we can argue that for $r_i=r_*$ the constituent branes become
massless. To get more information about this configuration we can now expand the result
for large values of the incision radius $r_i$. The coefficients of the metric components
in the longitudinal directions ($\mu,\ \nu$) can be interpreted as proportional to the
effective tension of the brane. So we see that:
\begin{equation}
\tau (r_i)=\frac{1}{2\kappa^2}\frac{H'}{H}=\frac{1}{2\kappa^2}\frac{ (p-7) +
\frac{Q^2}{(3-p)\rho r_i^{3-p}}}{1-\frac{Q^2}{\rho (3-p)(10-2p)r_i^{3-p}}}\frac{1}{r_i}\
\sim\
\frac{1}{2\kappa^2}\Bigg(\frac{p-7}{r_i}+\frac{Q^2}{(10-2p)\rho}\frac{1}{r_i^{4-p}}\Bigg)
\end{equation}
We can notice that this expression seems to indicate a configuration that does not
reproduce the behaviour of the source term in the equation of motion (\ref{H}).\\ Thus we
now imagine a slightly different configuration: we think that all the $N$ ordinary branes
are in the origin and the fractional branes alone form a shell at $r=r_i$. We can now
repeat the computation of the stress-energy tensor for this configuration and we get
always a vanishing contribution in the transverse directions but in the longitudinal ones
we have the following result:
\begin{equation}
\tau (r_i)\ \sim\ \frac{1}{2\kappa^2}\frac{M}{N}\frac{Q}{(10-2p)}\frac{1}{r_i^{4-p}}
\end{equation}
This result is now compatible with the source term which appear in (\ref{H}). This
expression for the effective tension is due
to the presence of $M$ fractional branes and it contains the small parameter $M/N$ that is typical for these expansions.\\
We get then the following solution:
\begin{eqnarray}
H_p(r) &=& h_s(r)+\Theta(r-r_i)[h_f(r)-h_f(r_i)]\\
\nonumber H'_p(r) &=& h'_s(r)+\Theta(r-r_i)h'_f(r)\\
\nonumber H''_p(r) &=& h''_s(r)+\Theta(r-r_i)h''_f(r)+\delta(r-r_i)h'_f(r)
\end{eqnarray}
where $$h_s(r)=\frac{\rho}{r^{7-p}}\ \ \ \mbox{and}\ \ \
h_f(r)=-\frac{Q^2}{(p-3)(10-2p)}\frac{1}{r^{10-2p}}$$ For the sake of clarity, we report
here the modified ansatz and the equations of motion for the case of a D$0$ fractional
brane, but everything continues to be easily generalizable to other $p$:
\begin{eqnarray}
&&\nonumber F_2=dt\wedge dH^{-1}\\
&&\nonumber B_2=\frac{Q}{p-3}\frac{\Theta(r-r_i)}{r^{3-p}}\ \omega_2\\
&&\nonumber \tilde{F}_4=H^{-1}\frac{Q\ \Theta(r-r_i)}{r^{4-p}} dt\wedge \ dr\wedge\omega_2\\
&&\nonumber d(\mbox{e}^{\phi/2}\ *\tilde{F}_4)=Q\delta(r-r_i)\\
&&\label{theta} d(\mbox{e}^{\frac{3}{2}\phi}\ *F_2)=g_s\mbox{e}^{\phi/2}H_3\wedge
*\tilde{F}_4\ =\
-\frac{Q^2}{r^{4-p}}\Theta(r-r_i)-\frac{Q^2}{p-3}\frac{\delta(r-r_i)}{r^{3-p}}
\end{eqnarray}
With the new solutions all the singular terms (see in particular the last equation) fit
perfectly. They are those which give rise to the junction conditions. 

Thus the scenario seems to be the following one. We have D$(p+2)$-branes at $r=r_i$ that
are wrapped around the 2-cycle which is Poincar\`e dual to $\omega_2$. According to the
form of their effective tension, it seems these objects extend in ($4-p$) of the
directions of the base of the cone. Moreover, these have a non trivial coupling with the
$B_2$ field, that becomes a source for fractional D$p$-branes potential for all $r\geq
r_i$ (notice the form of the source term in (\ref{theta})). \\ This is anyway suggested
also by the running of the fluxes:
\begin{equation} \Phi\ =\ \int_{{\mathcal{C}}_{9-p}} d(\mbox{e}^{\frac{3}{2}\phi}*F_2)\ =
\rho+\Big(\frac{1}{3-p}\frac{Q^2}{r^{3-p}}-\phi_*\Big)\Theta(r-r_*)
\end{equation}
where $\phi_*=\frac{1}{3-p}\frac{Q^2}{r_*^{3-p}}$. This behaviour seems indeed to indicate
that the fractional branes are present for all $r\geq r_i$ and that, at fixed $r$, they
extend in the ($4-p$) directions of the base of the cone where the shell at $r=r_i$
spreads.

In our discussion we have never fixed $r_i$, but we found a lower bound for it:
$$r_*^3=\frac{Q^2}{(7-p)(3-p)\rho},$$ where we argued the branes become tensionless. This
is very reminiscent of the enhan\c{c}on locus \cite{pol}, where we know the branes become
effectively massless. We think indeed it is now working the same mechanism, that seems
typical for fractional D-branes in various background \cite{bertolini,frau,pol}. In this
case it would be natural to identify the enhan\c{c}on radius $r_e$ with $r_*$.

\vskip0.5cm
\noindent{\bf The fractional D3-brane}

We spend now some words on the case of the fractional D3-brane \cite{conifold,KT,KS,buchel,non-extremal,restoration}. Because of its relevance to the gauge theory side, this case has already been
extensively studied from different perspectives: in \cite{KT}, Klebanov and Tseytlin found
a solution that exhibits a singularity in the IR. In a subsequent paper \cite{KS} Klebanov
and Strassler, getting insights from the strong dynamic of the infrared gauge theory,
argued that the naked singularity of the Klebanov-Tseytlin (KT) geometry should be
resolved via the deformation of the conifold. In \cite{buchel} instead this procedure has
been reversed, trying to make predictions on the field theory using string theory
computations. It was suggested that a non-extremal generalization of the KT solution may
have a regular Schwarzschild horizon ``cloaking'' the naked singularity. 
Non-extremal supergravity solutions translate in field theories at finite temperature. In this case, the dual field theory interpretation of this has been then supposed to be the restoration of chiral symmetry
at a finite temperature $T_C$ \cite{buchel}.

From the supergravity point of view, we are then in a particular  situation: at zero
temperature the theory exhibits chiral symmetry breaking \cite{KS}, while for temperature
above $T_C$ we have chiral symmetry restoration. So, the regular Schwarzschild horizon
should appear only for some finite Hawking temperature. As noticed in
\cite{buchel,non-extremal,restoration}, one implication is that, at temperatures below
$T_C$, there should be non-extremal generalizations of the Klebanov-Strassler solution
which are free of horizons, just like the extremal solutions. Particularly, the analysis
of \cite{buchel,non-extremal,restoration} suggests that gravitational backgrounds which
exhibit regular Schwarzschild horizon only above some critical non-extremality, should be
quite generic for backgrounds dual to gauge theories which undergo finite temperature
symmetry breaking phase transition. Nevertheless, they are unusual and, to my knowledge,
new from the supergravity point of view.

I argue that supergravity backgrounds of that type have to involve the enhan\c{c}on
mechanism in non-extremal backgrounds, so that a combination of the enhan\c{c}on shell
effect and of the horizon of the black hole, determines the appearance of regular
Schwarzschild horizons only in particular range of the parameters of the solutions.

\end{document}